\documentclass[twocolumn]{webofc}
\usepackage[varg]{txfonts}   % Web of Conferences font
\usepackage{bm}
\begin{document}
\title{Pushing the limits of excited-state $g$-factor measurements}

\author{\firstname{Andrew E.} \lastname{Stuchbery}\inst{1}\fnsep\thanks{\email{andrew.stuchbery@anu.edu.au}}
     \and
        \firstname{Brendan P.} \lastname{McCormick}\inst{1} \and
        \firstname{Timothy J.} \lastname{Gray}\inst{1} \and
        \firstname{Ben J.} \lastname{Coombes}\inst{1}
        }
%        \firstname{Third author} \lastname{Third author}\inst{3}\fnsep\thanks{\email{Mail address for last
%             author if necessary}}
%        % etc.
%}

\institute{Department of Nuclear Physics, Research School of Physics and Engineering, The Australian National University, Canberra ACT 2601, Australia
%\and
%           the second here
%\and
%           Last address
          }

\abstract{%
Current developments in excited-state $g$-factor measurements are discussed with an emphasis on cases where the experimental methodology is being extended into new regimes. The transient-field technique, the recoil in vacuum method, and moment measurements with LaBr$_3$ detectors are discussed.

%  Insert your english abstract here. 5 page limit
}
\maketitle
\section{Introduction}
\label{intro}

The magnetic moments of nuclear ground states provided important empirical evidence for the development of the nuclear shell model \cite{MariaMayer-PhysRev.78.16}. Today magnetic moment measurements on ground and excited nuclear states remain important observables to gain insights into nuclear structure -- they are sensitive to the single-particle structure of the quantum state, give insight into how the nucleus carries its angular momentum, and can distinguish single-particle versus collective contributions to the wavefunction.

This paper reviews some current developments in excited-state $g$-factor measurements.
 %wherein the experimental methodology is being developed into new regimes.
The transient-field technique (sect.~\ref{sec-1}), the recoil in vacuum method (sect.~\ref{sec-2}), and moment measurements with LaBr$_3$ detectors (sect.~\ref{sec-3}) are discussed.
As the gyromagnetic precession of the nucleus is the experimentally measured quantity, the following discussion generally refers to $g$~factors rather than magnetic moments: $g=\mu/I$ where $I$ is the nuclear spin and the moment $\mu$ is given in nuclear magnetons.

\section{Transient field measurements}
\label{sec-1}

\subsection{50 years of transient fields}

An intense hyperfine magnetic field called the transient field (TF) acts on the nuclei of ions moving swiftly within a magnetized ferromagnetic medium. The discovery of the TF \cite{bor68} will reach its 50th anniversary in 2018. While some $g$-factor measurements performed between 1968 and 1975 made use of the TF (e.g. \cite{EberhMg26}), it was not until after the 1975 discovery that the TF increases with the velocity of the moving ion \cite{EBERHARDT-ETF1975} that the method became widely used \cite{ben80}. It has continued in regular use ever since \cite{benczerkoller07}. The TF method gives the sign of the $g$~factor and is best suited for relative $g$-factor measurements on excited states with lifetimes in the picosecond range. The following subsection describes a contemporary measurement of $g(^{26}{\rm Mg};2^+)/g(^{24}{\rm Mg};2^+)$ by the high-velocity TF method \cite{Mg26.MCormick2018}.

\subsection{High-velocity transient-field method: $N=14$ subshell closure in $^{26}$Mg}

The $E(2^+_1)$ and $B(E2)$ systematics for $Z=12$ indicate a subshell closure at $N=14$, i.e. $^{26}$Mg: as $N$ increases from $^{22}_{12}$Mg$_{10}$ the $E(2^+)$ value spikes at $N=14$ and the $B(E2)$ value dips, indicative of a subshell closure. The expectation, then, is that the $2^+_1$ state of $^{26}$Mg should be dominated by proton excitations, giving $g(2^+_1)\sim+1$. Indeed, shell model calculations, using NuShellX \cite{NuShellX} and the USDB interactions \cite{Brown-New-USD,USDA-B_Obs}, predict $g(2^+_1)=+0.959$. Surprisingly, the currently adopted value is $g(2^+_1)=+0.50(13)$ \cite{A26,SpeidMg26}, half the expected value. %All experimental indicators of a shell or subshell closure should be consistent. The inconsistency of this $g$-factor measurement is therefore problematic.

The first $g(2^+_1$,~$^{26}$Mg) measurement by Eberhardt \textit{et al.} in 1974, using the thick foil transient-field method in which the excited $^{26}$Mg ions slowed and stopped in a magnetized iron host, found $g=+0.97(18)$ \cite{EberhMg26,ZalmTF}. Later, in 1981, Speidel \textit{et al.} \cite{SpeidMg26} argued that Eberhardt \textit{et al.} had incorrectly accounted for the static-field contribution, which came into effect after the ions came to rest in the iron host. Speidel \textit{et al.} made a new measurement using the thin-foil transient-field method, thereby excluding the static-field contribution, and obtained $g=+0.50(13)$, in agreement with Hartree-Fock calculations available at the time. This result, which implies near equal contributions from protons and neutrons, is currently listed as the adopted value in Nuclear Data Sheets \cite{A26}. As noted above, modern shell model calculations and single-particle arguments contend that the $N=14$ subshell closure should result in $g(2^+_1)$ being much more heavily influenced by the proton contribution than the currently adopted measurement indicates. Both Eberhardt \textit{et al.} and Speidel \textit{et al.} used $(\alpha,\alpha')$ reactions to excite and recoil $^{26}$Mg ions into an iron host. The recoil velocity was relatively low, $v/c$~$\sim$~1\%, and precession angles due to the transient field were very small, $\sim$1~mrad. These were challenging experiments.

%\begin{figure}[th]
%\centering
%\includegraphics[width=\columnwidth]{HVTFMg.pdf}
%\label{fig:HVTFsketch}
%\end{figure}

A high-velocity transient-field measurement \cite{StuchMgTF,Zn72-Fiori} on beams of $^{24,26}$Mg ions was performed. %Figure~\ref{fig:HVTFsketch} shows a sketch of the experimental arrangement.
Beams of 120-MeV $^{24}$Mg$^{8+}$ and $^{26}$Mg$^{8+}$ were provided by the ANU Heavy Ion Accelerator Facility. The beams were Coulomb excited on a 9.9~mg/cm$^2$ natural gadolinium target, which also served as the ferromagnetic layer for the transient-field precession effect. Precession angles an order of magnitude larger than the earlier works \cite{EberhMg26,SpeidMg26} were observed. Moreover, the same target was used with beam excitation to measure the ratio of $2^+_1$-state $g$ factors in $^{24}$Mg and $^{26}$Mg. Taking $g(2^+_1; ^{24}\rm{Mg}) = +0.538(13)$ \cite{Mg24RIV} gave $g(2^+_1; ^{26}\rm{Mg}) = +0.86(10)$.
The present $g$-factor measurement agrees with that of Eberhardt \textit{et al.} \cite{EberhMg26}, but with a reduced uncertainty. More details of the experiment and results are given in Ref.~\cite{Mg26.MCormick2018}.

\begin{figure}[h]
\centering
\includegraphics[width=\columnwidth]{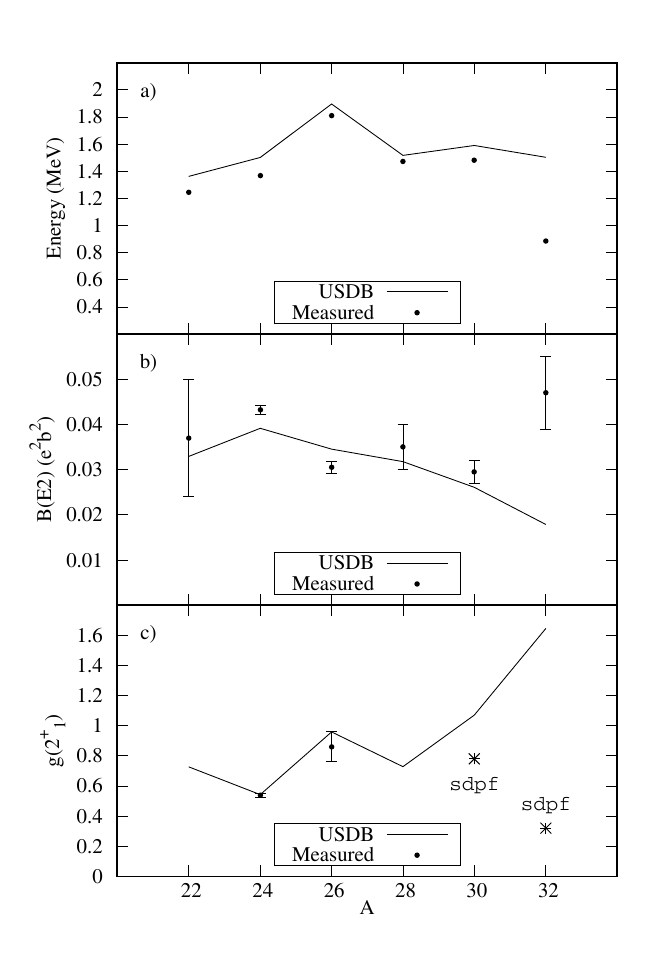}
\caption{ Comparison of USDB shell model calculations and experiment for the magnesium isotopes from $A=22$ to 32 a) $E(2^+_1)$ energies, b) $B(E2)$ rates, and c) $g$~factors. The theoretical $g$ factors for $^{30}$Mg and $^{32}$Mg in a more realistic $sdpf$ model space are shown as stars \cite{MCShellModel}.}
\label{fig:MGTF}       % Give a unique label
\end{figure}

The $E(2^+_1)$, $B(E2)$ and $g(2^+_1)$ systematics of the even-$A$ magnesium isotopes from $^{22}$Mg to $^{32}$Mg are shown in Fig.~\ref{fig:MGTF}. The peak in $E(2^+_1)$ and the dip in $B(E2)$ value at $^{26}$Mg together are indicative of the $\nu d_{5/2}$ subshell being filled. Shell-model calculations performed with NuShellX \cite{NuShellX} and the USDB interaction \cite{Brown-New-USD,USDA-B_Obs} are in agreement with most of the experimental data in Fig.~\ref{fig:MGTF}. Of course, as the Island of Inversion is approached toward $^{32}$Mg, the $sd$-shell model breaks down. More realistic predictions of $g$~factors in the $sdpf$ model space \cite{MCShellModel} are indicated in the lowest panel of Fig.~\ref{fig:MGTF}. To measure the magnitudes of these $g$~factors requires measurements on radioactive beams, for which the recoil in vacuum method has advantages over the TF method.

\section{Recoil in Vacuum}
\label{sec-2}

In general, although the RIV method gives only the magnitude of the $g$~factor, it has proven to give it more precisely than the transient-field method, particularly in the case of radioactive beam measurements where statistical precision is limited; compare Refs.~\cite{Allmond-Sn126RIV,Sn126TF}. The primary reason is that the transient-field method requires $\gamma$-ray detection at a few specific angles in the plane perpendicular to the direction of the applied magnetic field whereas the RIV method can take advantage of $\gamma$-ray detection over a much broader angular range. A second reason, applicable for
the time-dependent recoil in vacuum (TDRIV) method as applied recently to hydrogen-like $^{24}$Mg $2^+_1$ ions \cite{Mg24RIV}, is that the hyperfine interaction of the free ion in vacuum  can be calculated from first principles with very high accuracy. The case of simple atomic systems such as H-like and Li-like ions will now be discussed, followed by some remarks on the RIV measurements on complex ions, which still have to be calibrated empirically.

\subsection{RIV with H-like, Li-like and Na-like ions}

%For one-column wide figures use syntax of figure~\ref{fig-1}
\begin{figure}[t]
\centering
\includegraphics[width=\columnwidth]{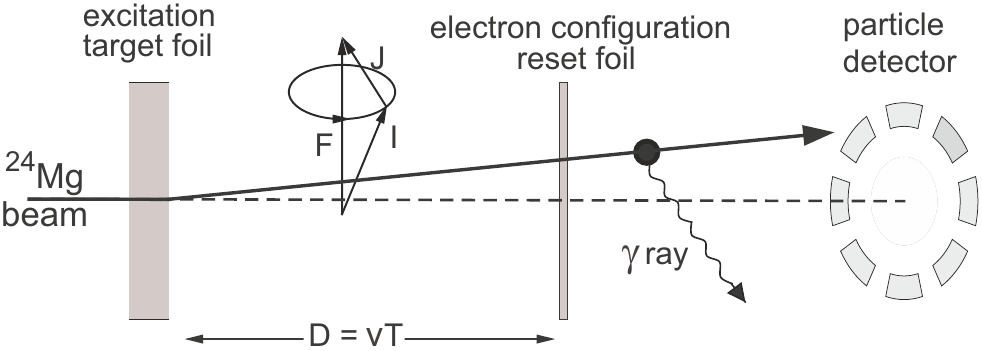}
\caption{Sketch of TDRIV experiment. The `stopper' of the traditional
plunger technique is replaced by a thin foil that resets the electron
configuration of H-like ions. The particle detector, with segmentation around the beam axis, is located
downstream of the $\gamma$-ray detectors.}
\label{fig:MgTDRIVfig1}       % Give a unique label
\end{figure}

The experimental method \cite{stu05ecr} used to measure $g(2^+_1)$ in $^{24}$Mg \cite{Mg24RIV} is illustrated in Fig.~\ref{fig:MgTDRIVfig1}. Excited nuclei emerge from a target foil as ions carrying one electron. The nuclear spin ${\bm I}$ is aligned by the Coulomb-excitation reaction whereas the atomic spin ${\bm J}$ is oriented randomly. The hyperfine interaction couples the atomic spin to the nuclear spin, and together they precess about the total ${\bm F} = {\bm I} + {\bm J}$ with a frequency proportional to the nuclear $g$~factor.
Thus the orientation of the nuclear spin is periodically reduced and restored during the flight through vacuum. As a consequence, the angular intensity pattern of the $\gamma$ rays emitted by the nuclei varies periodically, in step with the orientation of the nuclear spin. The nuclear precession frequency is determined by observing changes in the radiation pattern as the flight time is varied by changing the distance between the target and reset foils.

\begin{figure}[t]
\centering
\includegraphics[width=\columnwidth]{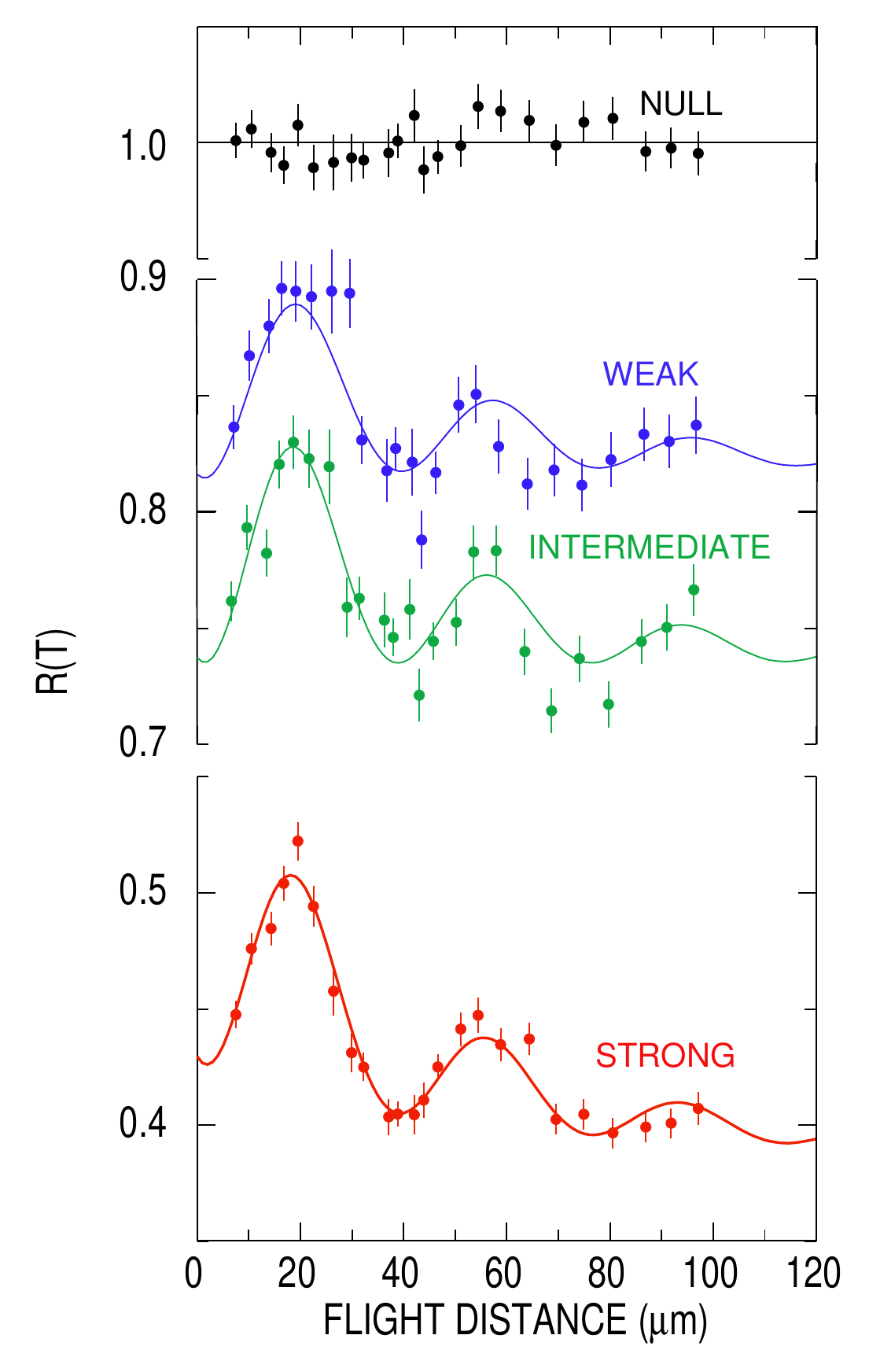}
\caption{ $R(T)$ ratio data showing the time-dependence of the radiation pattern and fits based on detailed modeling of the TDRIV experiment on $^{24}$Mg \cite{Mg24RIV}. The distance is the separation of target and reset foils ($22.4~\mu {\rm m} = 1$~ps flight time). $R(T)$ is a ratio of $\gamma$-ray intensities observed at different angles versus the flight time $T$ (or plunger separation). See \cite{Mg24RIV} for details. The frequency of the oscillation determines the $g$~factor whereas the amplitude affects the precision achieved. The rate of damping is determined by the nuclear level lifetime.}
\label{fig:Mg24Rt}       % Give a unique label
\end{figure}

Experimental data obtained at the ALTO facility at IPN Orsay showing the time dependence of the radiation pattern are shown in Fig.~\ref{fig:Mg24Rt} \cite{Mg24RIV}. Results of the fits to these $R(T)$ data having strong, intermediate and weak amplitude oscillations give $g= 0.538(13)$, 0.539(24) and 0.54(3), respectively, where the uncertainties are statistical only. The weighted average is $g= 0.538(11)$ (statistical error). Even the weakest amplitude oscillations yield a $g$~factor with a statistical uncertainty of better than 6\%. These data demonstrate that if the precession frequency can be observed in a TDRIV measurement, then the statistical precision is likely to exceed that of a transient-field $g$-factor measurement on the same state. %Moreover, the TDRIV method on simple ions can side step the uncertainties in effective hyperfine fields that pertain when the ions have complex atomic configurations, for which the net strength of the hyperfine field is an uncertain superposition of many components.

Unfortunately the TDRIV method can be applied to H-like ions only up to $Z \sim 20$ because the hyperfine frequency scales with $Z^3$ and the period becomes too short to measure; for $g=0.5$, the period of the oscillations is 3.1~ps at $Z=10$ and 0.38~ps at $Z=20$.

To apply this type of TDRIV method to higher-$Z$ nuclei requires the use of the weaker fields of shielded electrons, such as Li-like ions (2s electron) or Na-like ions (3s electron). The use of the method with Li-like or Na-like ions is under investigation. There are indications from measurements of charge-state distributions and integral recoil in vacuum measurements on relatively low-$Z$ ions that atomic ground states and low-excitation atomic states dominate the free-ion interactions \cite{Stuchbery2013HFI12}. However a Na-like ion, for example, has many more excited states than an H-like ion, with the potential to wash out the unique frequency of the atomic ground-state configuration. Along with the $^2S_{1/2}$ ground state the low-excitation $^2P_{1/2}$  and $^2P_{3/2}$ atomic states are likely to contribute. Even so, Lin et al. \cite{Lin1978}  have reported evidence of population of Na-like $^{41}$Ca ions in the $^2S_{1/2}$  state via the observation of a spin precession at the expected frequency in transverse decoupling experiments on the $15/2^+$ level with $\tau = 4.7$~ns.

\begin{figure}[t]
\centering
\includegraphics[width=\columnwidth]{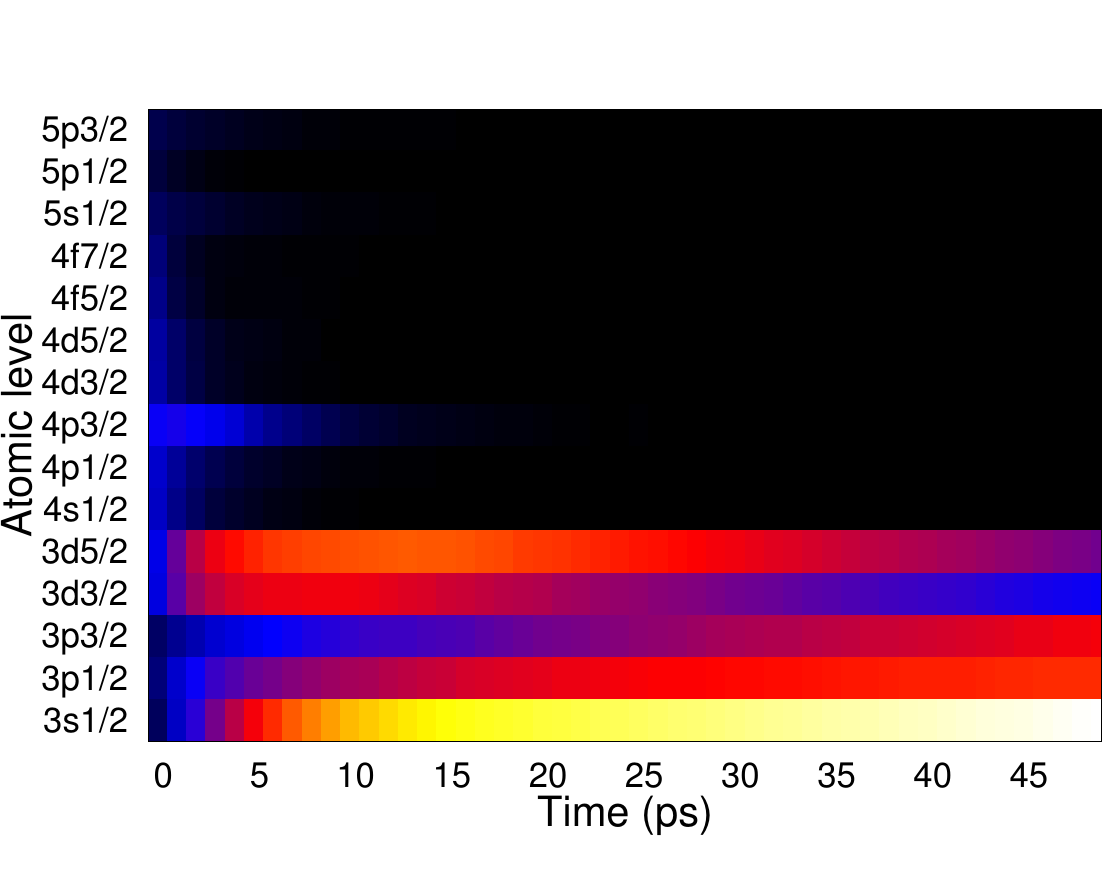}
\caption{ Evolution of the population of atomic states for Na-like Fe assuming an initial Gaussian distribution centred around 200 eV excitation energy with $\sigma = 100$ eV. Black represents least (no) population and white most.
}
\label{fig:Na-like}       % Give a unique label
\end{figure}

We have also begun to perform calculations with the GRASP2K atomic structure codes \cite{GRASP2K} to explore how the population of atomic states might evolve as ions recoil in vacuum.
The time evolution of excited atomic states of Na-like Fe assuming an initial Gaussian distribution of excited states is shown in Fig.~\ref{fig:Na-like}. Note that there is a prominent decay to the lower-excited states and  the ground state on the time-scale of several picoseconds. This calculation is schematic; at present it is unclear what initial population distribution is appropriate. However it does tend to confirm the empirical observations that atomic ground states and low-excitation atomic states dominate the free-ion interactions on the time scale of several picoseconds.

Attempts are underway to apply the TDRIV method to Na-like ions of $fp$ shell nuclei. Whether oscillations associated with the Na-like $^2S_{1/2}$ atomic ground state are observed or not, important information will be obtained to understand the free-ion hyperfine fields and develop their application to future $g$-factor measurements. If an identifiable oscillation frequency is clearly observed, it may provide an independent and reliable measure of the absolute $g$ factors in the $fp$ shell. Such measurements are very important because transient-field strengths in the $fp$ shell are somewhat uncertain due to a dearth of independently determined $g$~factors that can serve to calibrate the transient-field strength - see Ref.~\cite{PhysRevC.79.024303}.

\subsection{RIV with complex ions}

The time integral RIV technique on complex ions with $\sim 20$ - 30 electrons has proved to be a powerful method to measure the $g$~factors of excited states of neutron-rich nuclei produced as radioactive beams, particularly in the Sn and Te isotopes near the neutron-rich doubly magic nuclide $^{132}$Sn \cite{Stone.Te132g.PhysRevLett.94.192501,AES.TeRIV.PhysRevC.76.034307,AES.Te134.PhysRevC.88.051304,JMA.Sn124-8g.PhysRevC.87.054325,JMA.StableSn.PhysRevC.92.041303,Te136.PhysRevLett.118.092503,Te136g.PhysRevC.96.014321}.
One of the method's advantages is that the $g$~factor of the 2$^+_1$ state can be measured simultaneously with $B(E2; 0^+ \rightarrow 2^+)$ and $Q(2^+)$
\cite{AES.Te134.PhysRevC.88.051304,JMA.Sn124-8g.PhysRevC.87.054325,JMA.Sn124-8BE2.PhysRevC.84.061303,JMA.StableSn.PhysRevC.92.041303}.
Although the hyperfine interaction must be calibrated empirically, the RIV method has proven to give the magnitude of the $g$~factor more precisely \cite{Stone.Te132g.PhysRevLett.94.192501,AES.TeRIV.PhysRevC.76.034307,JMA.Sn124-8g.PhysRevC.87.054325} than the transient-field method
\cite{Te132TFg.BenczerKoller2008241,Sn126TF} A detailed description of calibration and analysis procedures has been included in the report on the case of $^{136}$Te \cite{Te136g.PhysRevC.96.014321}.

We are pushing the limits by developing this methodology for simultaneous measurements of $B(E2)$, $Q(2^+)$ and $g(2^+)$ in other regions of the nuclear chart. In particular, applications to few-nucleon $2^+_1$ states around $^{208}$Pb are of considerable interest. The shell structure in the neutron-rich $^{132}$Sn region can be compared with that in the vicinity of stable $^{208}$Pb \cite{Coraggio.PhysRevC.80.021305,Gargano09}. While the high-spin structure has been quite thoroughly studied experimentally around $^{208}$Pb, the electromagnetic properties of low-excitation, low-spin states associated with a few pairs of valence nucleons outside $^{208}$Pb have not. Thus direct comparisons of the related few-particle states around $^{132}$Sn and $^{208}$Pb are currently limited by the lack of experimental data on electromagnetic properties near $^{208}$Pb rather than near $^{132}$Sn. Measurements of the type described here on $^{210}$Pb, $^{210}$Po and $^{212}$Po would enable comparison with their equivalents, $^{134}$Sn, $^{134}$Te and $^{136}$Te. At present little is known about the strengths of the relevant free-ion hyperfine interactions near $Z=82$; it is important to determine the strength of the RIV interaction for ions with $v/c \sim 0.08$ and $\sim 30$ - 40 electrons recoiling in vacuum. The effective fields are expected to be much stronger than in the $^{132}$Sn region, so the $g$-factor measurements in the vicinity of $^{208}$Pb will need to control the interaction time of the nuclear moment with the electronic configuration by use of a plunger.

\section{Applications of LaBr$_3$ detectors}
\label{sec-3}

The development of Lanthanum Bromide (LaBr$_3$) detectors, which have excellent time resolution (typically {${\sim}300$~ps}) and energy resolution much superior to scintillators
such as NaI and BaF$_2$, provides an opportunity to perform perturbed angular
distribution $g$-factor measurements under new experimental
conditions. We have investigated the application of LaBr$_3$ detectors in beam to measure \mbox{${g}$
factors} of nuclear states with nanosecond lifetimes using static hyperfine magnetic
fields in magnetic hosts.

\begin{figure}[t]
\centering
\includegraphics[width=\columnwidth]{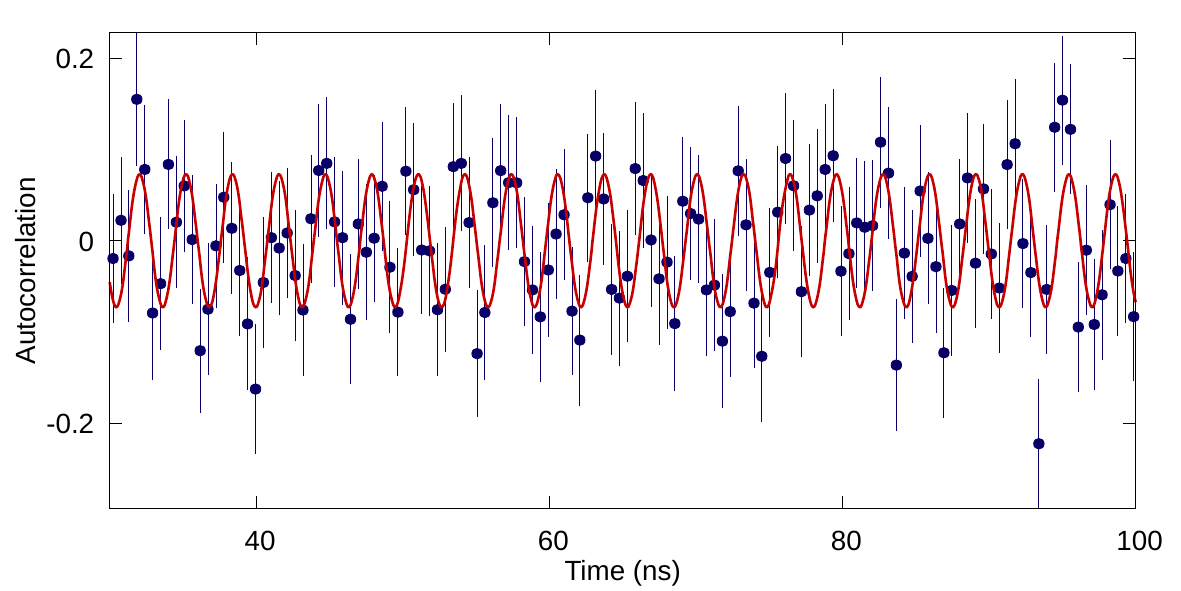}
\caption{Autocorrelation function for the 3432-keV transition below the 10$^+$ isomer of
$^{54}$Fe after implantation into nickel and showing the expected 3.3 ns oscillation period. The autocorrelation procedure was applied to the time-dependent perturbed angular distribution data, as described in Ref.~\cite{GG}, to make the oscillations visually more apparent.}
%}
\label{fig:Fe54auto}       % Give a unique label
\end{figure}

A preliminary experiment on $^{54}$Fe implanted into nickel was performed. The target consisted of 0.76 mg/cm$^2$ of $^{45}$Sc evaporated onto a nickel foil of thickness 2.44 mg/cm$^2$ that had previously been annealed at $\sim 790^{\circ}$C. The 10$^+$  isomer with $\tau = 525(10)$ ns and $g=+0.728(1)$ was populated by the $^{45}$Sc($^{12}$C, p2n)$^{54}$Fe reaction with 40-MeV $^{12}$C beams from the ANU 14UD Pelletron.
For the known $g$~factor and hyperfine field $B_{hf} = -27.00(3)$ T, the expected precession period is $\sim 3$ ns. This period approached the limit of the experimental setup because the time width of the beam-pulse was about 2 ns. The expected frequency was observed in the measured autocorrelation function for the 3432 transition as shown in Fig.~\ref{fig:Fe54auto}.
These results demonstrate that cases where the precession period is of the order of a few nanoseconds are
accessible for in-beam measurement.

%The autocorrelation $X(n)$ between two channels $k_1$ and $k_2$ is derived from the usual ratio of time spectra,
%$x_k=(N_k^{\uparrow}-N_k^{\downarrow})/(N_k^{\uparrow}+N_k^{\downarrow})$, where $N_k^{\uparrow (\downarrow)}$ is the number of counts in channel $k$ for field up (down), %and $X(n) \propto \sum_{k=k_1}^{k_2-n} {x_k x_{k+n}}/({k_2-k_1-n+1})$.

\begin{figure}[b]
\centering
\includegraphics[width=\columnwidth]{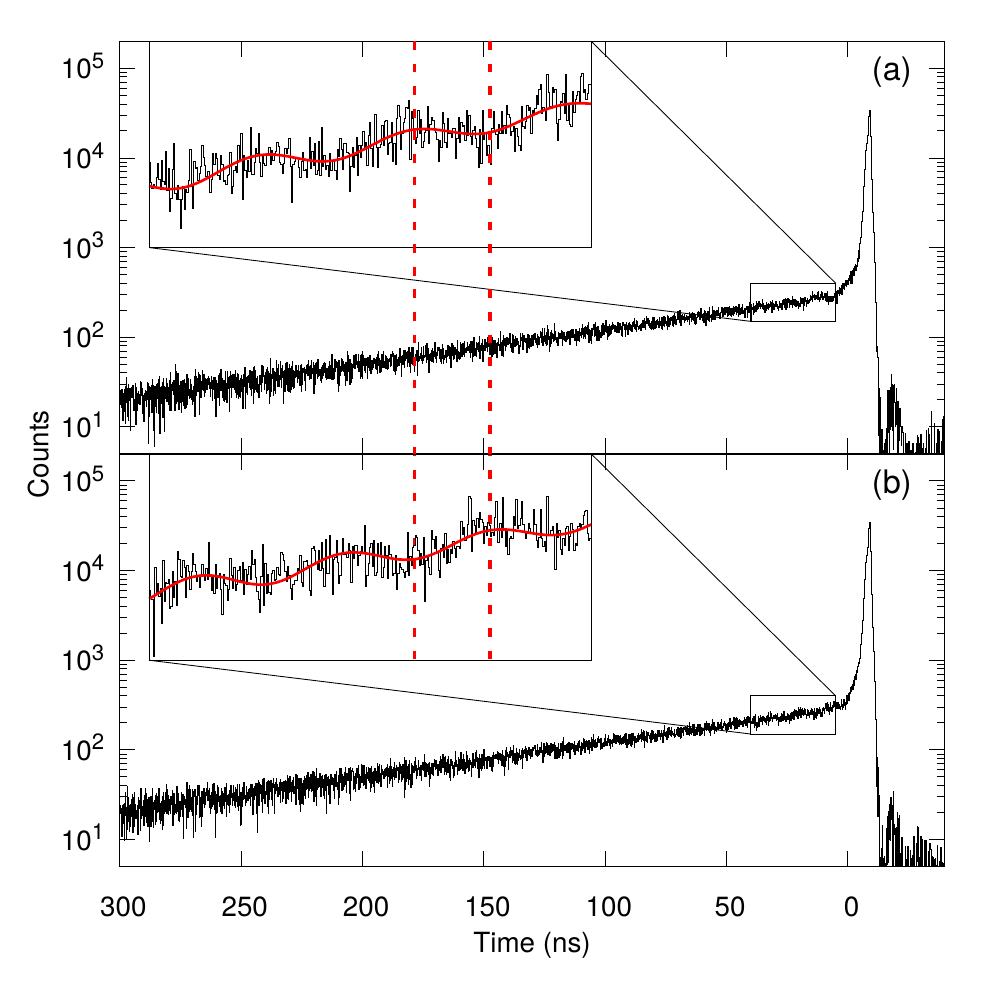}
\caption{Time spectra for the 640-keV transition depopulating the
    $I^\pi=\frac{11}{2}^-$ isomer in $^{107}$Cd. Time as displayed starts with the beam pulse and stops when a $\gamma$-ray is detected.
    Fits to guide the eye show the oscillations in the two spectra out of phase. The prompt
  peak is due to a prompt 632-keV transition in $^{106}$Cd that cannot be
  resolved from the 640-keV line by the LaBr$_3$ detectors.}
\label{fig:Cd107}       % Give a unique label
\end{figure}

The hyperfine field of Cd implanted into gadolinium was
then investigated.  The motivation was to reassess the $g$-factor measurement
on the \mbox{$I^\pi=10^{+}$} state in $^{110}$Cd which reported \mbox{$g(10^+)=-0.09(3)$}, at least a factor of two smaller than
\mbox{$g\approx-0.2$ to $-0.3$} that would be expected for a rather pure $(h_{\frac{11}{2}})^2$ neutron
configuration~\cite{Regan1995}.

The hyperfine fields at $^{107}$Cd ions implanted into gadolinium following the
$^{98}$Mo($^{12}$C, $3n$)$^{107}$Cd reaction were measured under similar conditions to the $^{110}$Cd $g$-factor measurement.
Examples of time spectra for the 640-keV transition depopulating the 11/2$^-$ isomer in $^{107}$Cd are shown in Fig.~\ref{fig:Cd107}.
The oscillations show close to the expected frequency, but their amplitude indicates that the fraction of nuclei on field-free sites is significant. The
consequence is that the effective hyperfine field for the integral perturbed angular distribution measurement of $g(10^+)$ in  $^{110}$Cd was much smaller than adopted in Ref.~\cite{Regan1995}. With a corrected effective hyperfine field a re-analysis of the data from Ref.~\cite{Regan1995} gives $g(10^+)=-0.29(16)$, consistent with that of the expected seniority-two $\nu h_{11/2}$
configuration. A full account of this work has been published \cite{TGray2017}.

\section{Conclusions}

Some current developments in excited-state $g$-factor measurements have been described. The transient-field method continues to give important results a half century after the transient field itself was discovered. The recoil in vacuum method has advanced greatly over the past decade, yet it has much potential for further development, particularly through time-dependent measurements. Moment measurements with LaBr$_3$ detectors and fast timing have opened up new opportunities for in-beam $g$-factor measurements on states that live a few nanoseconds.

\section*{Acknowledgements}
We thank our many colleagues who contributed to the work discussed here.
This research was supported in part by the Australian Research Council grant numbers DP140102986, DP140103317 and DP70101673. B.P.M. T.J.G. and B.J.C. acknowledge the support of the Australian Government Research Training Program. Support for the Heavy Ion Accelerator Facility operations through the Australian National Collaborative Research Infrastructure Strategy (NCRIS) program is acknowledged.

%
% BibTeX or Biber users please use (the style is already called in the class, ensure that the "woc.bst" style is in your local directory)

\bibliography{CGS16}
%\begin{thebibliography}{}
%Journal Author, Journal \textbf{Volume}, page numbers (year)
%Book Author, \textit{Book title} (Publisher, place, year) page numbers
% etc
%\end{thebibliography}

\end{document}